\documentclass[aps,prl,twocolumn,showpacs,amsmath,amssymb]{revtex4}
\usepackage{epsfig}
\usepackage{dcolumn}
\usepackage{bm}

\begin{document}

\title{Charge transfer statistics in symmetric fractional edge-state
Mach-Zehnder interferometer}

\author{V.V. Ponomarenko}
\affiliation{Center of Physics, University of Minho, Campus
Gualtar, 4710-057 Braga, Portugal}
\author{D.V. Averin}
\affiliation{Department of Physics and Astronomy,
University of Stony Brook, SUNY, Stony Brook, NY 11794}

\date{\today}


\begin{abstract}
We have studied the zero-temperature statistics of charge transfer
between the two edges of Quantum Hall liquids with filling factors
$\nu_{0,1}=1/(2 m_{0,1}+1)$ forming Mach-Zehnder interferometer. The
known Bethe ansatz solution for symmetric interferometer is used to
obtain the cumulant-generating function of charge at constant
voltage $V$ between the edges. Its low-$V$ behavior can be
interpreted in terms of electron tunneling, while its large-$V$
asymptotics reproduces the $m$-state dynamics ($m\equiv
1+m_{0}+m_{1}$) of quasiparticles with fractional (for $m>1$) charge
and statistics. We also analyze the transition region between
electrons and quasiparticles.
\end{abstract}


\maketitle

Mach-Zehnder interferometer (MZI) \cite{mz1,mz01,mz02}
based on the quantum Hall edge states in the regime of
the Fractional Quantum Hall effect (FQHE), together with
quantum antidots \cite{ant,any}, provides potentially
useful tool \cite{mz4,mz3,us} for the observation of
fractional statistics of FQHE quasiparticles. In
contrast to their fractional charge, which has been
confirmed in experiments \cite{ant,n1,n2}, in
particular, by measuring the Fano factor of a single
tunneling contact, there are no generally accepted
observations of the anyonic statistics of the
quasiparticles. One of the difficulties is that, in
typical experiments, the quasiparticles are produced at
the edges, i.e., they emerge as continuum of gapless
excitations described by a 1D field theory \cite{wen}.
In this theory, individual quasiparticles are tangible
only asymptotically in a special limit. In the
fractional edge-states MZI this limit occurs at
\emph{large} voltages, when the edge excitations are
quantized by the strong tunneling potential, and the MZI
Hamiltonian \cite{us} is dual to the usual Hamiltonian
of the electron tunneling.

In this work, we calculate the zero-temperature full
counting statistics \cite{fcs} of the charge transferred
between the two edges of Quantum Hall liquids forming
Mach-Zehnder interferometer (Fig.~1). The high-voltage
asymptotics of this statistics is explained by weak
tunneling of fractionally charged quasiparticles of
anyonic braiding statistics, which gives rise to the
$m$-state dynamics of the MZI as a result of successive
changes of the effective flux through it due to
quasiparticle tunneling. Since the low-voltage charge
transfer is still quantized in electrons, the found
counting statistics reflects the effect of the anyonic
phases on electron splitting into quasiparticles with
increasing voltage $V$ between the MZI edges. To
characterize this crossover conveniently, we study the
behavior of experimentally observable Fano factor of the
MZI at different voltages. We show, in particular, that
close to complete destructive interference in the
interferometer, the quasiparticle statistics ensures
that the MZI Fano factor retains its electronic value
$1$ for all voltages, including the quasiparticle
tunneling range. Away from this regime, the Fano factor
can reach the quasipaticle value $1/m$ at large
voltages, and exhibits non-monotonic voltage dependence
with a minimum indicating the crossover between
electrons and quasiparticles.

Our calculations start with the electronic model of MZI (Fig.~1)
with two single-mode edges of filling factors $\nu_{0,1}=1/(2
m_{0,1}+1)$, and $m_0 \geq m_1\geq 0$. In the standard bosonization
approach \cite{wen}, electron operator $\psi_l$ of the edge $l$ is
$\psi_l=(D/2\pi v_l)^{1/2} \xi_l \exp \{i[\phi_l(x,t)/ \sqrt{\nu_l}
+k_lx] \}$, where $\phi_l$ are the two chiral right-propagating
bosonic fields, which satisfy the usual commutation relations
$[\phi_l(x),\phi_p(0)] =i \pi \mbox{sgn}(x)\, \delta_{lp}$. The
Majorana fermions $\xi_l$ account for mutual statistics of electrons
in the opposite edges, and $D$ is a common energy cut-off of the
edge modes. The Fermi momenta $k_l$ correspond to the average
electron density in the edges, while the density fluctuations are:
$\rho_l(x,\tau)= (\sqrt{\nu_l}/2 \pi) \partial_x \phi_l(x, \tau)$.

\begin{figure}[htb]
\setlength{\unitlength}{1.0in}
\begin{picture}(3.,1.15)
\put(0.6,-0.15){\epsfxsize=1.9in \epsfbox{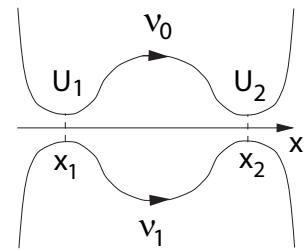}}
\end{picture}
\caption{Mach-Zehnder interferometer considered in this work: two
contacts with tunneling amplitudes $U_j$ formed at points $x_j$,
between single-mode edges with different filling factors $\nu_0$ and
$\nu_1$. The arrows show propagation direction of the edges.}
\end{figure}

In the symmetric interferometer with equal times of propagation
between the contacts along the two edges, it is convenient to
introduce the tunneling field $\phi(x)$:
\begin{equation}
\phi(x)\equiv [\sqrt{\nu_1}\phi_0(x)-\sqrt{\nu_0}\phi_1(x)]/
\sqrt{\nu_0 + \nu_1} \label{phi-} \, . \end{equation}
The Lagrangian for electron tunneling at points $x=x_{1,2}$ is
expressed then in terms of this field:
\begin{equation}
L_t = \sum_{j=1,2} (DU_j/\pi) \cos [\lambda \phi(x_j)+ \kappa_j ] \,
, \label{e2}
\end{equation}
where $U_j$ and $\kappa_j$ are the absolute values and the phases of
the dimensionless tunneling amplitudes, the products of the Majorana
fermions $\xi_1 \xi_2$ were omitted, since they cancel out in each
perturbative order due to charge conservation, and $\lambda \equiv
(\nu_0^{-1} + \nu_1^{-1})^{1/2}= \sqrt{2m}$. In the absence of
tunneling, the field $\phi$ is a free chiral right-propagating
bososnic field.  When $L_t$ is non-vanishing, $\phi$ undergoes
successive scattering at the points $x_{1,2}$ which breaks the
charge conservation and creates a tunneling current. The phases
$\kappa_j$ in (\ref{e2}) include the contributions from external
magnetic flux $\Phi_{ex}$ and from the average electron numbers
$N_{0,1}$ on the two sides of the interferometer: $\kappa_2 -
\kappa_1= 2\pi (\Phi_{ex} /\Phi_0)+ (N_0/\nu_0)-(N_1/\nu_1)]+
\mbox{const} \equiv -\kappa $. The voltage $V$ between the edges can
be introduced as a shift of the incoming field: $\phi_0-\sqrt{\nu_0}
Vt$, which translates into the following shift of $\phi \rightarrow
\phi -Vt/\lambda$.

A thermodynamic Bethe ansatz solution known \cite{ba} for one
point-contact with $\lambda^2=2m$ can be generalized \cite{us} to the
tunneling at two contacts by successive application of boundary
$S$-matrices \cite{za} to the bosonic field excitations: kinks,
antikinks, and breathers. For charge transport, only transitions
between the kinks and antikinks are important, and their $S$-matrices
are:
\[ {\cal S}^{\pm \pm}_{j,k}={(a k/T_{jB})^{m-1}
e^{i\alpha_{j,k}}\over 1+i(a k/T_{jB})^{m-1} } , {\cal S}^{-
+}_{j,k} = {e^{i(\alpha_{j,k} -\kappa_{j})}\over 1+i(a
k/T_{jB})^{m-1} }\, . \]
Here, the energy scales $T_{jB}$ characterize the
tunneling strength at the $j$th contact (explicit
relation to the electron tunneling amplitudes as follows
from the perturbative calculations \cite{us,ba} is given
next to Eq.~(\ref{lnP4})), and
\[ a\equiv 2 v \sqrt\pi\Gamma(1/[2(1-\nu)])/ [\nu
\Gamma(\nu/[2(1-\nu)])]\, . \]
At zero temperature, kinks fill out all available
``bulk'' states with momentum $k$ and distribution
$\rho(k)$ up to some momentum $A$ defined by the applied
voltage. Each kink with momentum $k$ undergoes
successive scattering at the two contacts described by
the product of the two boundary $S$-matrices,
independently of other quasiparticles. Therefore, the
large-time $t$ asymptotics of the cumulant-generating
function $\ln P(\xi )$ (logarithm of the Fourier
transform of the probability distribution of transferred
charge, which we measure in units of electron charge
$e=1$) is expressed as a sum of independent
contributions $\ln p(k,\xi)$ from individual momentum
states:
\begin{equation}
\ln P(\xi )=t \int_0^{A} dk \rho(k)\ln p(k,\xi) \, , \label{lnP1}
\end{equation}
where $p(k,\xi)$ is expressed as usual,
\begin{equation}
p(k,\xi)=1+\tau_C(k)(e^{i\xi}-1) \label{p1}
\end{equation}
through the total transition probability $\tau_C(k)$ of kink with
momentum $k$ into antikink. With parametrization $(T_{jB}/a)^2\equiv
\exp\{\theta_j/(m-1)\}$, it can be written as
\begin{equation}
\tau_C(k)=|(\hat{\cal S}_2\hat{\cal S}_1)^{-,+}|^2=B(
\tau(\theta_2,k)-\tau(\theta_1,k)) , \label{tauC1}
\end{equation}
through the transition probability in one contact
\[ \tau(\theta_j,k)=|\hat{\cal S}_j^{-,+}|^2=[1+k^{2(m-1)}
e^{-\theta_j}]^{-1} \, ,\]
and the factor characterizing interference:
\[ B(T_{jB},\kappa)=|T_{1B}^{m-1}+T_{2B}^{m-1}e^{i\kappa}
|^2/ [ T_{2B}^{2(m-1)}-T_{1B}^{2(m-1)}]\, . \]
Below, we take $\theta_2\ge\theta_1$, and write
$\theta_j$ as $\theta_{2,1}=\bar \theta\pm \Delta
\theta_0$ with dimensionless $\Delta \theta_0\ge 0$,
i.e., $\exp\{\Delta \theta_0 \}=(T_{2B}/T_{1B})^{m-1} $.

The aim of our subsequent derivation is to find the
cumulant-generating function $\ln P(\xi )$ in terms of the two
generating functions $\ln P_S$ for charge transfer in a single
contact found from the Bethe ansatz solution by Saleur and Weiss
\cite{sw}. This derivation does not need the explicit form of
$\rho(k)$ and A which can be found in \cite{ba}. Following the
approach for one contact, we first relate $\ln P(\xi )$ in
Eq.~(\ref{lnP1}) to the effective tunneling current. To do this, we
introduce the generalized tunneling probability
\begin{equation}
\tau_C(u,k)\equiv[1+(\tau_C^{-1}(k)-1)e^{-u}]^{-1} \, ,
\label{tauC2} \end{equation}
defined so that with the substitution $\tau_C(k)\rightarrow
\tau_C(u,k)$ in Eq.~(\ref{p1}), one has the following identity:
\[ (-i \partial_{\xi})^j \ln p(k,\xi)|_{\xi=0}=(\partial_u
)^{j-1}\tau_C(u,k)\, ,  \]
which shows that the cumulant-generating function is indeed related
to a tunneling current defined by $\tau_C(u,k)$:
\[ \partial_{i\xi} \ln P(u,\xi )/t =\int_0^{A} dk
\rho(k)\tau_C(u+i\xi,k) \equiv I(u+i\xi,V) \, . \]
Substituting Eq.~(\ref{tauC1}) into (\ref{tauC2}), one
finds $\tau_C(u,k)$ as difference of the single contact
transition probabilities $\tau(\bar \theta\pm
\Delta\theta(u),k)$, where $\Delta \theta(u)>0$ is
defined by the equation
\begin{equation}
\cosh\Delta\theta(u)=\cosh\Delta \theta_0 [1+
R(e^u-1)]\,, \label{theta1} \end{equation}
with $R\equiv B\tanh \Delta\theta_0$. Further substitution of
$\tau_C(u,k)$ into the definition of $I(u,V)$ results in the formula
\begin{equation}
I(u,V)=\partial_u\Delta \theta(u)\sum \pm I_{1/m}(\bar \theta\mp
\Delta\theta(u),V)\, , \label{IC}
\end{equation}
which gives the derivative of the generating function in
terms of the tunneling current $I_{1/m}(\theta,V)$ in
one single-point contact as found from the Bethe ansatz
solution \cite{ba}. Finally, integration of (\ref{IC})
over $u$ presents $\ln P(\xi)=\ln P(u,\xi)|_{u=0}$ as
the sum of two generating functions $\ln P_S$ for charge
transfer in a single contact:
\begin{equation}
\ln P(\xi )= \!  \sum_{j=1,2}  \ln P_S \big(V/T_{jB},
e^{(-1)^{j-1}(\Delta\theta(i\xi)-\Delta\theta_0)}\big) ,
\label{lnPs1} \end{equation}
defined \cite{sw} by the low- and high-voltage expansions:
\begin{eqnarray}
\frac{\ln P_S(s,e^{i\xi} )}{\sigma_0 Vt}= \sum_{n=1}^\infty {c_n(m)
\over m n} s^{2n({1\over \nu}-1)}(e^{i n \xi}-1), \; s<e^\Delta,
\nonumber\\ =i\nu \xi+ \sum_{n=1}^\infty {c_n(\nu)\over n}
s^{2n(\nu-1)}(e^{-i n\nu \xi}-1), \; s>e^\Delta, \;\;\label{lnPs2}
\\ c_n(\nu)=(-1)^{n+1 }{\Gamma(\nu n+1)\Gamma(3/2)\over
\Gamma(n+1)\Gamma(3/2+(\nu-1)n)}\, .\nonumber
\end{eqnarray}
Here $e^\Delta= (\sqrt{\nu})^{\nu/(1-\nu)}\sqrt{1-\nu}$, and
$\sigma_0$ is the conductance quantum. Although Eq.~(\ref{lnPs1})
seems to suggest that the charge transfer process in the MZI is
divided into two independent processes associated with two contacts
of the MZI, such a division is not complete. Each charge transfer in
Eq.~(\ref{lnPs2}) triggers multiple transfers in two contacts as
reflected in Eq.~(\ref{theta1}) for $\Delta \theta(u)$ which enters
Eq.~(\ref{lnPs1}).

At small voltages, $V<T_{1,2B}e^\Delta$, the two low-voltage
expansions in Eq.~(\ref{lnPs1}) can be combined as follows:
\begin{eqnarray}
\frac{\ln P(\xi )}{Vt\sigma_0} = \sum_{n=1}^\infty {c_n(m)\over m n}
\sum_j (V/T_{jB})^{2n(m-1)} \nonumber \\ \cdot ([\cosh (n \Delta
\theta (i\xi )) / \cosh (n \Delta\theta_0 ) ] -1 )\, .  \label{lnP4}
\end{eqnarray}
One can use here the standard expression for $\cosh(n
\Delta\theta(i\xi))$ as a polynomial of $\cosh(
\Delta\theta(i\xi))$, which is, according to
Eq.~(\ref{theta1}), a linear function of $e^{i\xi}$.
This shows that, in general, the $n$th term in
Eq.~(\ref{lnP4}) contains transfers of all numbers of
electrons up to $n$. When $T_{2B} \gg T_{1B}$, one finds
$R \rightarrow 1$ in Eq.~(\ref{theta1}), and the
transfer statistics (\ref{lnP4}) approaches that of one
point contact \cite{sw}, specifically the contact with
the largest electron tunneling amplitude $U_1$, related
to $T_{1B}$ by $T_{jB}= 2 D (\Gamma(m)/U_j) ^{1/(m-1)}$.
In this case, the $n$th term in the voltage expansion
series of the statistics describes the transfer of
exactly $n$ electrons. In the lowest order in $V$, the
MZI statistics (\ref{lnP4}) reduces to the Poisson
distribution, with the factor in front of $(e^{i\xi}
-1)$ equal to the average electron tunneling current
\cite{us}.

At large voltages, $V>T_{1,2B}e^\Delta$, steps similar to those in
the low-$V$ case, give:
\begin{eqnarray}
\frac{\ln P(\xi )}{Vt\sigma_0}=\sum_{n=1}^\infty {c_n(1/m)\over
n2^{n/m}} \big[\sum_j\big(\frac{T_{jB}}{V}\big)^{ 2(m-1)}
\big]^{\frac{n}{m}} \sum_\pm \big[1+R \nonumber \\
\cdot (z-1) \pm
([1+R(z-1)]^{2}\!\!\!-\cosh^{-2}\Delta\theta_0)^{1/2}
\big]^{{n\over m}}  \Big|^{z=e^{i\xi}}_{z=1}\!\! . \;\;
\label{lnP7}
\end{eqnarray}
Again, for $T_{2B} \gg T_{1B}$, the transfer statistics (\ref{lnP7})
coincides with that of one point contact \cite{sw}, but now with the
largest quasiparticle tunneling amplitude $W_2$ related to $T_{2B}$,
by $T_{jB} =  2 m D (W_j/\Gamma(1/m))^{m/(m-1)}$. The $n$th term in
the expansion (\ref{lnP7}) corresponds to transfer of in general
fractional charge $n/m$ by $n$-quasiparticle. However, the
large-voltage asymptotics (the $n=1$ term) of the generating
function (\ref{lnP7}) can not be interpreted as a Poisson process of
the lowest-order tunneling of individual quasiparticles of charge
$1/m$. The reason for this is that the tunneling of each
quasiparticle changes the interference phase by the statistical
contribution $2\pi/m$ \cite{usprb05} and therefore the tunneling
rate of the next quasiparticle.

The appropriate description of such a dynamics of $m$ phase states
with different tunneling rates $\gamma_l$ should be based, as usual,
on the kinetic equation for probabilities to find MZI in one of the
states, $l=0,\, ...,\, m-1$. The equation can be written
conveniently \cite{feldman2} in terms of the $m$ dimensional
vector-function $Q_l(z,t)=\sum_n d_{l,n} z^{n/m}$, where
$z=e^{i\xi}$, and $d_{l,n}$ are the probabilities to have MZI in
state $l$ and $n$ tunneled quasiparticles:
\begin{equation}
\partial_t Q_l(z,t)=\sum_k M(z)_{l,k} Q_k(z,t) \, ,\label{kineq}
\end{equation}
where the transition matrix $M(z)_{l,k}$ is
\begin{equation}
M(z)_{l,k}=-\gamma_l\delta_{l,k}+\gamma_{k}z^{1/m}\delta_{l,k-1} \,
,\label{M} \end{equation}
with the Kronecker symbol $\delta_{l,k}$ defined modulo $m$. In what
follows, we show that the leading large-$V$ term of the generating
function (\ref{lnP7}) gives the same quasiparticle tunneling
statistics as the kinetic equation (\ref{kineq}). By doing this, we
also extend this equation to the situation of different edge filling
factors, when, as shown below, the quasiparticle exchange statistics
creates an additional shift $(m-1)\pi/m$ of the common interference
phase.

We note first that the large-$V$ limit of Eq.~(\ref{lnPs1}) is
expressed through the quasiparticle amplitudes $W_j$ as
\begin{equation}
\ln P(z)=tK[2W_1W_2\cosh( \Delta\theta(z)/m)-W^2_1-W^2_2] \,
,\label{lnP8}
\end{equation}
where $K \equiv \sigma_0V(2mD/V)^{2(m-1)/m} c_1(1/m)/\Gamma^2(1/m)$.
On the other hand, kinetic equation (\ref{kineq}) gives  \cite{bn}
the same generating function as $\ln P(z)=t\Lambda$, where $\Lambda$
is the maximum eigenvalue of the transfer matrix (\ref{M}), i.e., the
solution that goes to zero at $z \to 1$ of the equation
\begin{equation}
\prod_l (\gamma_l+\Lambda)-z\prod_l \gamma_l=0\, . \label{Lambda1}
\end{equation}

To see the equivalence of the two results we look for a
general solution of Eq.~(\ref{Lambda1}) with the
tunneling rates $\gamma_l$ in the form
$\gamma_l=K|W_1+W_2\exp\{i\phi_l\}|^2$, where
$\phi_l=\phi+2\pi l/m$ with some unknown $\phi$.
Substituting these expressions into (\ref{Lambda1}) and
using the two identities,
\begin{equation}
x^m-1=\prod_l (x-e^{i2\pi l/m}) \label{identity}
\end{equation}

\vspace*{-4ex}

\[ 2^{m-1}\!\!\prod_l (\cos\phi_l\!+\!\cosh (\frac{
\Delta\theta}{m})) =[\cosh\Delta\theta \!-\!(-)^m\!\cos(m\phi)] \,
,\]
where the second follows from the first, one confirms that
$\Lambda=\ln P(z)/t$ (\ref{lnP8}) solves Eq.~(\ref{Lambda1}) if
\begin{equation}
\cosh\Delta\theta=(-)^m \cos(m\phi)+{z\over 2} \prod_l
{\gamma_l\over K W_1 W_2}\, .\nonumber
\end{equation}
Calculating the RHS of this equation again with the help of
Eq.~(\ref{identity}), one can see that it gives precisely the
definition of $\cosh\Delta\theta(z)$ (\ref{theta1}), if $\phi$ is
taken as
\begin{equation}
\phi=\kappa/m +(m-1)\pi/m \, . \label{fi}
\end{equation}
This equation shows that besides the statistical shift
$2\pi/m$ of the interference phase between the MZI
states $l$ and $l+1$, the quasiparticle exchange
statistics also shifts the common phase $\phi$ from the
externally-defined value $\kappa/m$. This deviation of
$\phi$ from $\kappa/m$ is not important if $m$ is odd.
However, for even $m$ this shift changes the
``spectrum'' of the interference phases $\phi_l$. This
ensures that there is no shift of the interference
pattern of the tunnel current, e.g., its Fano factor
(see below), between electron and quasiparticle regimes.

So far, we have established the dynamics of electron and
quasiparticle tunneling that underlies the respective small- and
large-voltage limits of the charge transfer statistics. Now we focus
on the full range of the voltage dependence of the cumulants
associated with this statistics to study the crossover between the
two asymptotic regimes. The cumulants are defined by Eq.~(\ref{IC})
as
\begin{equation}
\langle N^j(t)\rangle_c/t=\partial_u^{j-1}I(u,V)|_{u=0}\, .
\label{cum}
\end{equation}
The first cumulant, $I \equiv I(0,V)= \langle N(t)\rangle/t$, is the
average tunneling current $I$, while the second cumulant is the
spectral density of the current fluctuations at zero frequency
$S_I(0) \equiv  \langle N^2(t)\rangle_c/t$, which at zero
temperature reflects the shot noise associated with charge transfer.
Equation (\ref{cum}) gives for this cumulant:
\begin{equation}
S_I(0) =(1-B\coth \Delta \theta_0)I -B^2\sum_{j=1,2}
\partial_\theta I_{1/m}(\theta_j,V)\, . \label{I2}
\end{equation}

The Fano factor $F=S_I(0)/I$ reflects both the charge
and statistics of the tunneling excitations and
illustrates the transition between the electron and
quasiparticle regimes in MZI. In the low-voltage limit,
$F=1$ as a result of the regular Poisson electron
tunneling. In the quasiparticle, large-voltage, limit,
we have from Eq.~(\ref{I2}),
\[ F=1-{|W_{1}^{m}+W_{2}^{m}e^{i\kappa}|^2
\over W_{2}^{2m}-W_{1}^{2m}} \big\{{W_{2}^{2m}+W_{1}^{2m} \over
W_{2}^{2m}-W_{1}^{2m}}-{1\over m}{W_{2}^{2}+W_{1}^{2} \over
W_{2}^{2}-W_{1}^{2}}\big\} \, . \]
This Fano factor corresponds to the dynamics of quasiparticle
tunneling as described by the kinetic equation (\ref{kineq}).
Because of the complex nature of this dynamics characterized by $m$
different rates $\gamma_l$, $F$ is not equal simply to the
quasiparticle charge $1/m$ but varies as a function of parameters
between $1/m$ and $1$. This variation is illustrated most clearly by
the case $m=2$, when
\[ F=1- |W_1^2+W_2^2e^{i\kappa}|^2/[2(
W_1^2+W_2^2)^2] \, . \]
For $\kappa=0$, the phase shift (\ref{fi}) is such that the two
quasiparticle tunneling rates coincide, $\gamma_1=\gamma_2$,
regardless of the ratio of the amplitudes $W_j$. Then $F=1/2$
demonstrating explicitly the quasiparticle charge $1/2$. Even in the
quasiparticle regime, $F$ reduces to electron value 1, if $W_1\simeq
W_2$ and $\kappa=\pi$. In this case, one of the rates $\gamma$ is
much smaller than the other, and on the large time scale set by the
smaller rate, the quasiparticles effectively tunnel together,
restoring $F$ back to 1.

For larger $m$, the Fano factor $F$ can be calculated numerically.
Figure 2 shows $F$ for $m=3$, e.g., for tunneling between the two
$\nu=1/3$ edges. The curves are shown for different degrees of
asymmetry between the two contacts, and for two values of the
interference phase: $\kappa=0$, and $\kappa=\pi$. Similarly to
$m=2$, destructive interference drives $F$ to its electron value 1
at all voltages. The transition region between electrons and
quasiparticles is indicated by the minimum of the Fano factor, with
$F$ decreasing even below the ``pure'' quasiparticle value $1/m$.
For symmetric junctions, such small value of $F$ can be understood
as a result of screening of the charge transfer in one contact by
the other contact.

\begin{figure}[htb]
\setlength{\unitlength}{1.0in}
\begin{picture}(3.,1.95)
\put(0.15,-0.15){\epsfxsize=2.7in \epsfbox{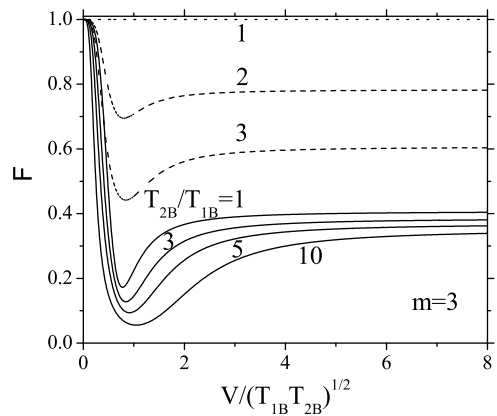}}
\end{picture}
\caption{The zero-temperature Fano factor $F$ of the Mach-Zehnder
interferometer formed by two $\nu=1/3$ edges, as a function of the
bias voltage $V$ for different degrees of asymmetry, $T_{1B}/T_{2B}$,
of the two contacts. The solid/dashed lines show the two values of
the interference phase: $\kappa=0$ and $\kappa=\pi$, respectively.
The curves illustrate the transition from the electron regime $F=1$
at small voltages to the quasiparticle $m$-state tunneling dynamics
at large voltages. } \end{figure}

In conclusion, starting from the exact solution of the
tunneling model of symmetric Mach-Zender interferometer
in the FQHE regime, we have calculated the statistics of
the charge transfer between interferometer edges. The
statistics shows the transition from electron tunneling
at low voltages to tunneling of anyonic quasiparticles
of fractional charge $e/m$ and statistical angle
$2\pi/m$ at large voltages. Electron tunneling is
characterized by the standard Poisson process. Dynamics
of quasiparticle tunneling is more complicated and
reflects the existence of $m$ degenerate phase states of
the interferometer. Quasiparticle braiding statistics
shifts the common interference phase by $(m-1)\pi/m$ in
these states, and also changes the phase by $2\pi/m$
from state to state. Crossover between electrons and
quasiparticles manifests itself as minimum of the Fano
factor of the tunneling current.

V.P. acknowledges support of the ESF Science Program
INSTANS, and Grant No. PTDC/FIS/64926/2006.

\end{document}